\documentclass[aps,prd,twocolumn,groupedaddress,nofootinbib,showkeys]%
{revtex4-1}

\usepackage[T1]{fontenc}
\usepackage{enumerate}
\usepackage{bm}
\usepackage{amsmath}
\usepackage{amssymb}
\usepackage{graphicx}
\newcommand{\R}{\Bbb R}

\newcommand{\bk}{{\bm k}}
\newcommand{\bl}{{\bm l}}

\newcommand{\bL}{{\bm L}}

\begin{document}

\title{Quantum measurement and fuzzy dark matter}

\author{Adam D. Helfer}

\email[]{helfera@missouri.edu}
\affiliation{Department of Mathematics and Department of Physics \& Astronomy,
University of Missouri,
Columbia, MO 65211, U.S.A.}

\date{\today}

\begin{abstract}
It has been suggested that dark matter is a superfluid of particles whose masses are on the rough order of $10^{-22}$ eV.
Since the occupation numbers are huge, the state is coherent, and the 
speeds typical of orbital velocities in halos, it has generally been assumed that a classical effective non-relativistic treatment is adequate.  However, the Compton wavelength would be $\sim 1\, {\rm pc}$, 
and around the Compton scale concerns about some aspects of quantum measurement theory, known in principle but not quantitatively significant in previous cases, become pronounced.
I estimate here the stress--energy operator, averaged over a few Compton wavelengths; a rough but useful approximation has a remarkably simple form.
Conventional quantum measurement theory gives physically unacceptable results:  a thought-experiment to measure the stress--energy is described which would involve only a modest apparatus but would
excite particles in the observation volume to relativistic energies; these particles would escape the Galaxy, and there would be a substantial violation of energy conservation.  Related foundational questions come up:  the meaning of measurements of observables with continuous spectra, and the problem of predicting when measurements occur.  
The effective classical theory of fuzzy dark matter is not affected; however, the underlying quantum theory cannot be regarded as satisfactory without resolving these issues.
But we may interpret the results more broadly.
The macroscopic Compton scale amplifies inadequacies of measurement theory which have not previously seemed pressing.
\end{abstract}



\maketitle

\section{Introduction}
\label{sec:intro}

Known forms of matter and conventional gravitation theory cannot explain the dynamics of galaxies and clusters, nor their formation.  While it remains possible that modifications of gravity will sort out these issues, at present most attention is given to the idea that some form of `dark matter' will resolve them.
An especially interesting proposal is that this should consist of ultralight particles (of mass perhaps $10^{-22}\, {\rm eV}$); see especially Hu et al. \cite{HBG2000}, and Hui et al. \cite{HOTW2017} for a recent detailed treatment of many aspects.  Such models are known as `fuzzy', `wave' or `quantum' dark matter.  Their hypothesis is that the particles are actually in a superfluid state, with streams moving at typical orbital speeds within their halos.  The de Broglie wavelengths are of the order of $1\, {\rm kpc}$, and this delocalization allows fuzzy dark matter to avoid small-scale problems which occur for other dark matter candidates.

Although motivated quantum-theoretically, 
the treatment of dark matter for these 
purposes has been entirely classical \cite{HOTW2017}.  Indeed, the perspective in fuzzy dark matter investigations is that, because one has a superfluid with a huge (on astrophysical scales) number density of particles in a coherent state, one can regard the quantum physics as `integrated out' and restrict attention to an effective classical theory.   
The expectation of any normal-ordered field observable in a coherent state will be the corresponding effective classical quantity, so the dispersions of observables are proportional to commutators and typically very small; therefore one might think that quantum effects will be macroscopically negligible.

\subsection*{The Compton scale}

There is, however, reason to hesitate.  The Compton length is macroscopic, and indeed of astrophysical size, on the order of $1\, {\rm pc}$.  The behavior of quantum fields changes markedly at this scale; in particular, a knot of unresolved issues in quantum measurement, present in principle at all scales, become prominent. 

It is a basic feature of relativistic quantum field theory that, at and below the Compton scale, measurements necessarily implicate relativistic modes.  Even
a state which is initially non-relativistic will have, after a 
Compton-scale 
measurement, relativistic excitations.  
A well-known example of this is that attempts to localize a particle at or below this scale will give its momentum relativistic components, and also give a significant probability of creating additional pairs of particles.
In such situations, the measurement has clearly not conserved the energy of the particle state; indeed the discrepancy is of the same order as the particle's original energy.

This problem --- that quantum measurements do not generally respect conservation laws --- has been investigated at least since Wigner~\cite{Wigner1952}.  While it has never been definitively resolved, it has not usually been considered to be worrisome, since in laboratory measurements it is plausible that any failure of conservation in the observed system could somehow be absorbed by the far larger measuring apparatus.  
However, it is important to appreciate that this does {\em not} mean that our {\em present} treatment of measurement is consistent with conservation laws.
Arguments like Wigner's support a view that the failures can be made {\em small} for `large' apparatuses, but also that the failures {\em cannot} be strictly eliminated, within conventional quantum theory~\cite{Wigner1952,AJR}.

But in the case of fuzzy dark matter, because the Compton scale is so large, we will see that the ratio of the sizes of the measuring apparatus to the observed system can be reversed.
It is the energies of the measuring devices which may be relatively small, with the dark energy content large ($\sim 10^{-2}\, M_\odot\, {\rm pc}^{-3}$ in our vicinity).  So the Wigner-type arguments do not apply, and there 
might be substantial conflicts between measurements and conservation;
we will see that this is indeed the case.  

I have focused the discussion on conservation laws, and these are of critical importance, but there are other difficulties with quantum measurement as well.  One of these is the question of what it means to measure an observable with a continuous spectrum.  Perhaps the most basic problem is to give an objective criterion for when a measurement occurs.  At the moment, we know of no quantity we can compute which would (for example) tell us how likely it is that, given the state of a system, a particular observable would be measured within a particular interval.\footnote{`Objective reduction' theories aim to address questions like this \cite{Leggett2002}.}  We will be led to look at these problems, too, by thinking about fuzzy dark matter around the Compton scale.

One resolution would be simply that ultralight-mass fields are impossible.  But we should remember that the function of the macroscopic Compton length is really to amplify problems which in principle are present at all scales.  It seems better to view these results as an opportunity to address the inadequacies of quantum measurement theory.

\subsection*{The stress--energy}

Dark matter interacts almost exclusively gravitationally, so its observables must be derived from the stress--energy.
In this paper, 
I will estimate the effects of measuring this for fuzzy dark matter on scales of a few Compton lengths.  There are two main reasons for this choice.  One is that many of the effects are proportional to the size of the observation volume, tending to favor larger choices (although not ones much beyond the Compton scale).  
But also it turns out that, if we are willing to settle for rough approximations, we can get remarkably simple formulas for the stress--energy in this case.

The expectation-values of the stress--energy are, of course, simply what would be calculated from the effective classical theory.  It is the states resulting from the measurements --- the projections of the original state onto the observables' eigenspaces --- which are problematic.    
These 
states turn out to contain essentially uniform distributions of excited modes, up to the wavenumber set by the measuring scale.
Most of these modes are relativistic, making the resulting state very different from the original superfluid.
If such a measurement occurred, it would populate a substantial fraction of the modes in the observational volume with relativistic particles, which would escape the Galaxy.  Fuzzy dark matter would be unstable against quantum measurements.\footnote{Extant observations of Galactic orbital velocities essentially probe the gravitational potential, and are too coarse to give
measurements of (its gradient) the stress--energy on parsec scales.}

This is a disquieting result, and one should ask whether it could be rejected by some known physical considerations.  (Could the parsec-scale measurements be unfeasible, even in principle --- perhaps one would need an enormous device whose own gravitational field would wash out the effect sought?)  However, I shall describe a thought-experiment, somewhat similar to the ideas of Khmelnitsky and Rubakov~\cite{KR2014}, which would effect the measurements with a modest apparatus.  So it appears the problem does lie with applying conventional quantum measurement theory to fields with ultralight masses.

One attempt to resolve this might be to maintain that, since none of the problematic observations have yet been made, there is no conflict.  
`Cavalier' is a mild word to describe this view, as it would mean that the stability of the Galaxy is hostage against such observations.  And this points up
a further difficulty.  What we have seen is that conventional theory predicts that 
small, relatively inconsequential, apparatuses can be used to measure the stress--energy, resulting in unacceptable changes to the quantum state.  If this is indeed the case, 
we should wonder whether other physical processes, not requiring human interventions, could also result in measurements.  
This leads back to the fundamental problem I mentioned earlier, that we do not have a theory of when quantum measurements occur.
In the case of fuzzy dark matter, this cannot be dismissed, because
we require a theory which explains Galactic dynamics for the past few Gy.  We would need a good argument that few problematic measurements, whether effected by  humans or other processes, could have taken place. 

Two other points are worth noting here.

The question of what it means to measure operators with continuous spectra has long been discussed.  
In the regime we consider, the pressure appears as a generator of squeezes; its spectrum is continuous, and its failure to commute with the energy is pronounced.  
I will show that a natural attempt 
to measure the pressure with even modest resolution would lead to states with energies wholly out of scale with the other quantities in the problem.  
This effect can be interpreted in terms of squeeze operators, with no reference to fuzzy dark matter or quantum field theory, and is of correspondingly general interest:  Attempts to measure the generators of squeezes would lead to states with very large expected occupation numbers, and (at least for the models investigated here) there is a limitation on the accuracy of those measurements if we require finite expected occupation numbers.

The second point is that, because dark matter interacts almost exclusively gravitationally, its observables must be derived from the stress--energy and must correspond to geometric effects --- in other words, the effects are quantum-gravitational, although we are very far from the Planck scale.  We will see this explicitly when we consider thought-experiments; it requires some care to control the potential dependence of the apparatus on the quantum-gravitational state.

This paper will not attempt to provide any solution to the problems of quantum measurement theory; its goal is rather to describe circumstances in which they are presented unavoidably, and from a new perspective.

\subsection*{Literature}

I have already mentioned Wigner's seminal paper; further references to this line of thought can be found in ref.~\cite{AJR}, and are often cast in terms of the WAY (Wigner--Araki--Yanase) Theorem.  Ref. \cite{AJR} also provides an entr\'ee into measurement theory for operators with continuous spectra.
The bulk of all of this work builds on von Neumann's~\cite{vN1955} general formalism, and that work contains a number of important insights (although just what von Neumann considered the physical interpretation of measurement itself remains debatable\footnote{He emphasizes the necessity for measurement to be represented by projection of the state vector, and to be distinguished from unitary evolution.  But
his most direct statement about what it {\em is} seems to hold that it is beyond physics:  `... it is inherently correct that the measurement or the related process of the subjective perception is a new entity relative to the physical environment and is not reducible to the latter.' Page 418 in \cite{vN1955}.}).  
All of these works involve some technicalities in the their formulations, and, as one is dealing with foundational questions, it is important to sort through these in connecting the mathematics to the physics.

Concerns, related to the present ones, about measurements and energetics
in connection with black-hole evaporation appear in \cite{ADH2004a,ADH2011,BMS2016}.

\subsection*{Organization}

Here is the plan of the paper.  Section II derives a rough approximation for the stress--energy, valid for estimates to within a factor of a few of its leading terms (in an expansion in relativistic effects) for its averages over a few Compton lengths.  We find the leading contributions are an average energy density $\overline\rho$ and and average isotropic pressure $\overline{P}$, both expressible in terms of a single annihilation operator $A$.  
Sections~III and IV analyze measurements of $\overline\rho$ and $\overline{P}$; Section~IV considers the measurement of operators with continuous spectrum taking into account finite-energy constraints.  
Section~V gives thought-experiments for measuring the stress--energy, and Section~VI discussion.  There is an appendix, giving the eigenstates of the squeeze generator and related computations.

{\em Notation and conventions.}  Conventions for quantum field theory are those of Schweber~\cite{Schweber1961}; for general relativity, those of Penrose and Rindler~\cite{PR1984}.  The metric signature is $+{}-{}-{}-$.  Factors of the speed of light and Planck's constant are not given explicitly; Newton's constant is $G$.

\section{The stress--energy in the moderate sector}

We will be concerned with the quantum fields over scales less than the gravitational radius of curvature ($\sim 1\, {\rm Mpc}$ for a galaxy).  On these, to good approximation, the field behaves as a special-relativistic Klein--Gordon one.  It can be written in terms of annihilation and creation operators $a(\bk )$, $a^*(\bk )$ as
\begin{equation}
\phi (x)=2^{-1/2}(2\pi )^{-3/2}\int\frac{d^3\bk}{\sqrt{E(\bk )}}
 \left( e^{-ik_ax^a}a(\bk )+\textrm{H. c.}\right)\, ,
\end{equation}
where $k^a$ is the wave four-vector, with spatial part $\bk$ and temporal component
$E(\bk )=\sqrt{m^2+\|\bk\|^2}$.
The corresponding renormalized stress--energy operator is
\begin{widetext}
\begin{eqnarray}\label{seeq}
 T_{ab}=2^{-1}(2\pi )^{-3}\int \frac{d^3\bk}{\sqrt{E(\bk)}}\frac{d^3\bl}{\sqrt{E(\bl)}}
  &&\left[ e^{-i(k_a+l_a)x^a}a(\bk)a(\bl)(-k_al_b+(1/2)\eta _{ab} (m^2+k\cdot l))
    +\textrm{Hermitian conjugate}  \right.\qquad\nonumber\\
    &&\left.
    +2e^{-i(k_a-l_a)x^a}a^*(\bl )a(\bk )(k_al_b-(1/2)\eta _{ab}((k\cdot l) -m^2)
    \right]\, .
\end{eqnarray}    
\end{widetext}
This is complicated, but if we are willing to settle for a rough approximation we can get remarkably simple formulas.

The stress--energy is not really well-defined pointwise; it must be averaged. 
Let the length scale for the averaging be $L$, which we will take to be a few times the Compton scale.  The corresponding wavenumber will be $\Lambda =2\pi /L$.  I will call the modes with wavenumbers below this scale {\em moderate}; they include the subrelativistic modes, but also have ones with appreciable, although not dominant, relativistic contributions.

Consider the effects of the averaging on each
of $e^{-i(k_a+l_a)x^a}a(\bk )a(\bl )$, $e^{i(k_a+l_a)x^a}a^*(\bk )a^*(\bl )$, $e^{-i(k_a-l_a)x^a}a^*(\bl )a(\bk )$.  When both wave-vectors are moderate, there is no significant interference.  On the other hand, there will be interference unless at least one of $\bk\pm\bl$ is moderate.  This means that if one $\bk$, $\bl$ is larger than about $\Lambda$, the other must be, too, if the term is not suppressed.\footnote{We will suppose the temporal averaging is less than $L$.  Longer temporal averages tend to suppress the $e^{-i(k_a+l_a)x^a}a(\bk )a(\bl )$, $e^{i(k_a+l_a)x^a}a^*(\bk )a^*(\bl )$ contributions.}

Very roughly, then, the contributions to the averaged stress--energy which are not suppressed by interference are of two sorts:  where both the wave-vectors are moderate, or where neither is (and this latter case is restricted by $\|\bk\pm\bl\|$ being moderate).  In this sense, the stress--energy respects a division of the modes into two sectors:  those which are moderate, and those which are not.  
While the division is not sharp, it will be adequate for our purposes, because fuzzy dark matter is supposed to be deeply subrelativistic.  The argument just given does show that any couplings of such modes to ones outside the moderate sector are suppressed.

It will therefore be enough for us to consider the moderate sector $\|\bk\|,\, \|\bl\|\leq\Lambda$.  This is the same as introducing a cutoff $\Lambda$ in the integral (\ref{seeq}).
We may then make the zeroth-order approximations
$k_a=mt_a =l_a$ and
$E(\bk ) =E(\bl ) =m$ for the c-number factors
in eq. (\ref{seeq}).
The averaged stress--energy is then
\begin{widetext}
\begin{eqnarray}\label{avT}
{\overline T}_{ab}&=& 2^{-1}(2\pi )^{-3} m
\left[-
  \left( \int _{\|\bk\|\leq\Lambda} d^3\bk  e^{-ik_ax^a}a(\bk)\right)
 \left( \int _{\|\bl\|\leq\Lambda} d^3\bl  e^{-il_ax^a}a(\bl )\right)(t_at_b-\eta _{ab})
  +\textrm{h.c.}   \right.\nonumber\\
  &&\left.
  + 2\left( \int _{\bk\|\leq\Lambda} d^3\bl  e^{il_ax^a}a^*(\bl)\right)
   \left( \int _{\bk\|\leq\Lambda} d^3\bk  e^{-ik_ax^a}a(\bk )\right)t_at_b\right]
   +\cdots
\, ,
\end{eqnarray}
\end{widetext}
where the ellipsis indicates terms which are not purely in the sector and also terms beyond zeroth order in the tensorial dependence.  The terms shown should, conservatively, give the purely moderate-sector effects to within a factor of a few.

We can simplify this.  Let 
\begin{equation}
  A =(4\pi\Lambda ^3/3)^{-1/2}\int _{\|\bk\|\leq \Lambda} d^3\bk
 \, e^{-ik_ax^a}a(\bk)\, .
\end{equation}
(This operator depends on the point $x$ at which the stress--energy is measured.)
Then $A$ is an annihilation operator with the standard discrete normalization
\begin{equation}
  [A,A^*]=1\, ,
\end{equation}    
and we may write
\begin{equation} 
{\overline T}_{ab} ={\overline\rho} t_at_b+{\overline P}(t_at_b-\eta _{ab})
+\cdots\, ,
\end{equation}
where
\begin{equation}\label{rhq}
\overline\rho = m(2\pi)^{-3} (4\pi\Lambda ^3/3) A^*A
\end{equation} 
and
\begin{equation}\label{peq}
\overline{P} =-2^{-1} m(2\pi )^{-3}(4\pi\Lambda ^3/3) \left( A^2+(A^*)^2\right)
\end{equation}
are the leading averaged energy density and averaged pressure operators in the moderate sector.  (In this approximation, the pressure is isotropic and there is no momentum density.)

We see that the averaged energy density has the same formal structure as a harmonic oscillator; its eigenvalues are simply $m\Lambda ^3 n$, where $n=0,1,2,\ldots$.  
However, in our case
the expected occupation numbers are $\sim 10^{83}$, so the spacing is very fine on the scales of interest.
The averaged pressure is the generator of a squeeze operator; its spectrum is the real line (with multiplicity two).

\section{Measurements of $\overline\rho$}

Let the quantum state be a coherent state corresponding to the complex classical state $\phi _{\rm cl}$ (of purely positive frequency), that is
\begin{equation}\label{cohst}
  |\Psi\rangle = \exp\left[ -(1/2)\omega (\overline{{\phi _{\rm cl}}},\phi _{\rm cl}) +
     \omega(\phi ,\phi _{\rm cl})\right] |0\rangle\, ,
\end{equation}     
where $\omega (f,g) =(i/2)\int (f\partial _t g-(\partial _t f)g)\, d^3x$.
We will assume that $\phi _{\rm cl}$ has no modes with relativistic wavenumbers.  

When a measurement of $\overline\rho$ is made, the state (\ref{cohst}) is projected into an eigenspace of this operator.  In order to work this out, let us introduce the operators
\begin{equation}
  b(\bk ) =a(\bk ) -(4\pi \Lambda ^3/3)^{-1/2} e^{ik_ax^a}\chi _\bk  A \, ,
\end{equation}
end where 
\begin{equation}
\chi _\bk =\begin{cases} 1&\text{if } \|\bk\|\leq\Lambda\\
    0 &\text{otherwise.}\end{cases}
\end{equation}    
Then $[b(\bk ),b(\bl ) ] =0$ and $[A,b^*(\bk ) ] =0$  and
\begin{equation}
[b(\bk ),b^*(\bl ) ] =\delta (\bk ,\bl )-(4\pi\Lambda ^3/3)^{-1}\chi_\bk \chi _\bl e^{-i(l_a-k_a)x^a}\, .
\end{equation}
(The operators $b(\bk )$ are linearly dependent, since $\int  \chi _\bk e^{-ik_ax^a} b(\bk ) \, d^3\bk=0$.)  We also set
\begin{eqnarray}
  \psi (y) &=&2^{-1/2}(2\pi )^{-3/2} \int \frac{d^3\bk}{\sqrt{E(\bk )}}
    e^{-ik_ay^a} b(\bk )  +\text{H. c.} 
    \, ;\qquad
\end{eqnarray}
then
\begin{widetext}
\begin{eqnarray}
|\Psi\rangle &=&
  \exp\left[ -(1/2)\omega(\overline{{\phi _{\rm cl}}} ,\phi _{\rm cl}) 
    +\omega (\psi ,\phi _{\rm cl}) 
     + 2^{1/2}(2\pi)^{3/2}(4\pi\Lambda ^3/3m )^{-1/2}\phi _{\rm cl}(x)A^*
      \right] |0\rangle\nonumber\\
      &=& \exp\left[ -(1/2)\omega(\overline{{\phi _{\rm cl}}} ,\phi _{\rm cl}) 
    +\omega(\psi ,\phi _{\rm cl})\right]   
    \sum _{n=0}^\infty (n!)^{-1}\left( 2^{1/2}(2\pi)^{3/2}(4\pi\Lambda ^3/3m )^{-1/2}\phi _{\rm cl}(x)A^*\right)^n|0\rangle
      \, .\qquad
\end{eqnarray}      
This sum gives us a spectral resolution of $|\Psi\rangle$ for the operator $\overline\rho$; the $n^{\rm th}$ term 
\begin{eqnarray}\label{rhoeig}
 \Pi _n|\Psi\rangle
   &=& \exp\left[ -(1/2)\omega (\overline{{\phi _{\rm cl}}} ,\phi _{\rm cl})
    +\omega(\psi ,\phi _{\rm cl})\right]   
     (n!)^{-1}\left( 2^{1/2}(2\pi)^{3/2}(4\pi\Lambda ^3/3m )^{-1/2}\phi _{\rm cl}(x)A^*\right)^n|0\rangle  
\end{eqnarray}
gives us the projection to the $n^{\rm th}$ eigenstate.  
It is a Poisson distribution.

\end{widetext}

The state (\ref{rhoeig}) resulting from the measurement differs markedly from an acceptable fuzzy dark matter state.  
It is the factor $(A^*)^n|0\rangle$ which is problematic, for the operator $A^*$ creates an essentially even distribution of particle modes over the mass shell up to the wavenumber $\Lambda$, and these modes mostly have relativistic wave-vectors.  Thus rather than the $n$ deeply subrelativistic particles a fuzzy dark matter state would have in this volume, we have an $n$-particle state with what I have called moderate momenta, that is, with appreciable but not dominant relativistic contributions.  In particular, these particles would not be gravitationally bound; they would escape their host galaxy and supercluster.

The argument of the previous paragraph is schematic, for two reasons.  First, to really assess the mode-content of the state (\ref{rhoeig}) one must know the action of the number operators $a^*(\bk )a(\bk )$ on it (not just $A^*A$).
Second, one would like to know what the effects of measuring $\overline\rho$ with finite resolution are, since its 
eigenvalues are so finely spaced.

It turns out that the operator $\overline\rho$ is simple enough that we can do the analysis explicitly, and the 
conclusions do not change as long as that resolution is even modestly below the expected value $\langle\Psi |\overline\rho |\Psi\rangle$.  Suppose, for example, the state has been measured and found to be in the subspace with $N_1\leq A^*A \leq N_2$.  
Let $|\Psi _{N_1N_2}\rangle$ be the resulting normalized state.
Then, assuming that $\|\bk\|$ is larger than the wavenumbers contributing to $\phi _{\rm cl}$, a straightforward if slightly lengthy calculation shows
\begin{eqnarray}\label{aven}
&&(4\pi\Lambda ^3/3) \langle\Psi _{N_1N_2} |a^*(\bk )a(\bk )|\Psi_{N_1N_2}\rangle \nonumber\\
  &&\qquad =\frac{N_{\rm cl} ^{N_2+1}/(N_2!) +N_{\rm cl} ^{N_1}/(N_1!)}{
    \sum _{n=N_1}^{N_2} N_{\rm cl} ^n/(n!)}\, ,\qquad
\end{eqnarray}
where
\begin{equation}
  N_{\rm cl}=(4\pi (\Lambda /2\pi )^3/3)^{-1} m|\phi _{\rm cl}|^2
\end{equation}
is of the order of
what the effective classical theory would give for the number of particles in the observation volume (before the measurement).\footnote{Because of the rough approximations made in specifying the volume and averaging, one cannot sharply say how many particles the effective theory would place there.}

If $N_2<N_{\rm cl}$, the higher powers in (\ref{aven}) are the dominant terms.  If, for instance $N_1/N_{\rm cl}$ and $N_2/N_{\rm cl}<1$ are considered fixed, for large $N_{\rm cl}$ the expression (\ref{aven}) is close to $N_{\rm cl}$.
Apart from the deeply sub-relativistic modes contributing to $\phi _{\rm cl}$, then, the effect of the measurement is to uniformly populate the mass-shell up to the cut-off, with as many excitations as there were `effective particles' in the observation volume.  The preponderance of these excitations will be relativistic (and so in particular energy has not been conserved), and the corresponding particles would escape the Galaxy.

A word about measurements of $\overline\rho$ over different volumes versus measurements of the total Hamiltonian is in order.  These do {\em not} commute, and so the total Hamiltonian cannot be observed by adding observations of $\overline\rho$ for different volumes.  Closely related to this, the need to `smear' the stress--energy in order to get a well-defined operator means we cannot really speak of the the energy in a sharply demarcated domain.  This is accommodated in our calculation by the cut-off in wavenumbers. 

\section{Measurements of $\overline{P}$}

Both the behavior and the analysis of the average pressure are different from those of the average density, because $\overline{P}$ is
a generator of squeezes rather than a number operator. 
The main result will be that measuring $\overline{P}$ results in states with energy contents far worse than those for $\overline\rho$. 
Partly, we will have to sort out physical questions about the measurement of an operator with continuous spectrum; also the terms $A^2$, $(A^*)^2$ in $\overline{P}$ are much more strongly non-commutative with the total energy than is $A^* A$ in $\overline\rho$.

It will be helpful to write $\overline{P} =-(2\pi )^{-3} (4\pi\Lambda ^3/3) mS$, where
\begin{equation}
  S=(1/2)\left( A^2+(A^*)^2\right)
\end{equation}   
is a dimensionless squeeze generator.  Using the formulas above, it is straightforward to compute
\begin{equation}
  \langle\Psi | S|\Psi\rangle =(2\pi )^2(4\pi\Lambda ^3/3m)^{-1}
    (\phi _{\rm cl}^2 +{\overline\phi}_{\rm cl}^2)\, .
\end{equation}    
This will typically be of the same order
as 
$\langle\Psi|A^*A|\Psi\rangle$, although $S$ may have either sign.    
The
eigenstates of $S$ are computed in the Appendix; we denote them $|\gamma_{\nu \pm}\rangle$ where $\nu\in\R$ and the sign is related to a parity.  The normalization is $\langle\gamma_{\nu \sigma}|\gamma_{\acute\nu \acute\sigma}\rangle =\delta (\nu -\acute\nu ) \delta _{\sigma ,\acute\sigma}$.

Let us first consider the measurement of an operator with continuous spectrum in a more familiar context.
Suppose we try to measure the position of a particle on a line in ordinary (non-relativistic) quantum mechanics.  The position operator 
cannot itself be 
measured, in the sense that there are no normalizable eigenvectors.  We must imagine dividing the line up into bins of finite size, and measuring which of these bins the particle is in.

In and of itself, such a theory of measurement is acceptable, but it runs into trouble when we start investigating the energetics of the situation.  For instance, if we  have a free particle, and then measure which bin it is in, 
the resulting wave function --- the original wave function projected to lie within the bin --- necessarily has components with arbitrarily high energies.  (A function with compact support has a Fourier transform which is analytic, and so has contributions from arbitrarily high wave-numbers.)

It seems reasonable that a real measurement of a particle can communicate only a finite amount of energy to it, and in this sense no real measurement 
which has bins of precisely defined positions and extents can exist.  {\em A real device, allowed only a finite amount of energy, will measure an approximation to the position, even allowing for binning.}  
The actual operator such a finite-energy device measures cannot have eigenstates which are strictly demarcated in space.  

Just these sorts of issues come up for measurements of 
$\overline{P}$, although of the problematic localization is in its eigenvalues (not physical position).  
One finds that $\langle\gamma _{\acute\nu \acute\sigma}|a^*(\bk ) a(\bk )|\gamma _{\nu \sigma}\rangle$, which would give us a measure of the energy-content of its eigenstates,
exists only in a very weak sense.  It is not even strictly speaking a distribution\footnote{With a change of variables, one can see it contains Fourier transforms of exponentials.}; this means it is only defined when integrated against a restricted class of very smooth test functions of the eigenvalues $\mu$, $\nu$.  The smoothness restrictions mean that only measurements which are sufficiently smeared in the eigenvalues are admissible.  Arbitrarily high-resolution measurements of $\overline{P}$ are energetically unacceptable.

A real measurement cannot therefore not return even a mathematically exact binning of eigenstates 
$|\gamma _{\nu \pm}\rangle$ of $\overline{P}$; it will rather produce some
smeared approximation to such a state.
We shall suppose for simplicity that 
this is a Lorentzian smearing 
with width $\Delta\nu$.  (Similar results hold for smearing by hyperbolic cosines.)

In the Appendix, such Lorentzian states, denoted
 $|\xi _{\nu \pm }\rangle$, are given explicitly.  The projection of the state $|\Psi\rangle$ to this eigenspace will be
 \begin{equation}
 |\Psi _{\nu\pm }\rangle =
   (\text{normalization}) e^{\omega(\psi ,\phi _{\rm cl}) } |\xi _{\nu \pm}\rangle\, .
\end{equation}   
The expectation $\langle\Psi _{\nu\pm}| a^*(\bk )a(\bk )|\Psi _{\nu\pm}\rangle$ (assuming $\|\bk\|$ is greater than the wavenumbers occurring in $\phi _{\rm cl}$) can be found in the Appendix.  It is complicated, but only two details about it are relevant here.  The first is that the expectation converges only for $\Delta\nu >2$; this is the quantitative consequence of the need for smearing described above.  If we were interested in states with low occupation numbers and small values of the squeeze, it would be a serious restriction; here, however, we typically expect $\nu$ to be enormous and $\Delta\nu$ must itself realistically be taken quite large.

The second, much more serious --- indeed, damning --- result of the computation is that, even for modest resolutions the energy-contents 
are grossly unacceptable.  Taking $|\nu|$ to be around its nominal generic value $\langle \Psi| A^* A|\Psi \rangle$ and assuming $\Delta\nu /|\nu|\lesssim 1$, we find
\begin{equation}
 (4\pi\Lambda ^3/3)\langle\Psi _{\nu\pm}| a^*(\bk )a(\bk )|\Psi _{\nu\pm}\rangle
   \simeq \nu ^2/2\, .
\end{equation}
In other words, the mass shell is uniformly populated (up to the cutoff), but by what is typically of the order of {\em the square of the number of classically expected particles}.  This is of the order of the square of the corresponding result for $\overline\rho /m$.

Finally, much of this analysis, although not the final conclusion, applies to squeezes in general (and does not depend on properties of fuzzy dark matter or quantum field theory).  The computation of the generators' spectra and eigenstates holds generally.  While I gave the discussion of energetic concerns in terms of the operators $a^*(\bk )a(\bk )$, the same sort of analysis applies for $A^* A$, and so the same sorts of difficulties in measuring $S$ with finite energies (taken as finite values of the number operator $A^* A$) hold.  The expectation $\langle\gamma _{\acute\nu  \acute\sigma} |A^* A|\gamma _{\nu \sigma}\rangle$ exists only very weakly.  For the smearings I have tried, only resolutions with $\Delta \nu >2$ appear admissible, and the expectation is of order $\nu ^2$.
These difficulties are related to the scale-free character of the eigenstates of $S$.

\section{A timing model}

It is instructive to consider a thought-experiment aimed at measuring components of the stress--energy.  The first issue we will encounter is that we must, for self-consistency, admit certain quantum-gravitational behavior.  While interesting conceptually, in general this leads to models which are very difficult to control.  However, we will find one which can be simply analyzed.  The most important conclusion will be that it does seem possible in principle, with modest apparatus, to measure the components of the stress--energy discussed earlier.  

The model to be investigated here is conceptually similar to the idea behind gravitational-wave detection via pulsar timing arrays.  Let us work in linearized gravity, and consider two freely falling bodies, an emitter which gives off signals at regular intervals according to its proper time $\tau _{\rm e}$, and a receiver which detects them at its proper time $\tau _{\rm r}$.  Thus the receiver records a function $\tau _{\rm e}(\tau _{\rm r})$.  We also assume that to zeroth order (in the metric perturbation) there is no velocity between their world-lines.  Then we have
\begin{equation}\label{timeq}
  \frac{d^2\tau _{\rm e}}{d\tau _{\rm r}^2} =\int R_{abcd} t^aL^bt^cL^d\, ds\, ,
\end{equation}
where $R_{abcd}$ is the linearized Riemann curvature, and the integral is taken over the null geodesic from the emitter to the receiver, whose tangent is $L^a$, with affine parameter $s$~\cite{ADH2013}.

Equation (\ref{timeq}) shows that the times of arrival of signals from the emitter will be affected by the intervening gravitational field.  It is just this effect which is the basis for pulsar timing array searches for gravitational waves, and in fact Khmelnitsky and Rubakov \cite{KR2014} suggested, based on a classical analysis, that the oscillatory character of fuzzy dark matter's pressure might allow its detection by such arrays.

The precise suggestion of Khmelnitsky and Rubakov would not be helpful for investigating the effects of interest here, for two reasons.  First,
the distances between pulsars, and between pulsars and the Earth, are many Compton-lengths.  Second, the geometry in their scheme does not give a very clean link to the stress--energy; the formulas are too complicated to be a good first model of the quantum effects to investigate. 
To see this and deal with it, I will push the analysis of this geometry a bit further, and then modify it to a spherically symmetric one.

It is straightforward to compute the curvature from the stress--energy, by working in the de Donder gauge; one finds the linearized metric perturbation is
\begin{equation}\label{Gfeq}
  h_{ab}=-16\pi G (\partial ^b\partial _b)^{-1} \left( T_{ac}-(1/2) T_{ac}\right)\, ,
\end{equation}  
where we take the retarded solution.
In the moderate sector, this gives
\begin{widetext}
\begin{eqnarray}
h_{ab}&=&8\pi G (2\pi )^{-3}m\int_{\|\bk\| ,\, \|\bl\|\leq\Lambda} d^3\bk d^3\bl
  \left[ (2m)^{-2}e^{-2im t}a(\bk )a(\bl ) (-(1/2)\eta _{ab}-t_at_b)
    +\text{Hermitian conjugate}\right.\nonumber\\
    &&\left.
    +2e^{-i(k_a-l_a)x^a} [(k_a-l_a)(k^a-l^a)]^{-1} a^*(\bl )a(\bk ) (t_at_b -(1/2)\eta _{ab})\right]\, .
\end{eqnarray}    
This follows from eq. (\ref{Gfeq}) by cutting off the creation and annihilation operators to modes with wave-numbers $\leq\Lambda$ and (therefore) approximating $E({\bf k})\simeq E({\bf l})\simeq m$.  Any theory of quantum gravity which has Einstein's equation valid as an operator equality at the linearized level will give the same result in this sector.  (Two further comments are in order.  First,
in principle, one should add an infinitesimal timelike imaginary part to the factors $k_a-l_a$ in the denominator, but the singularity turns out to be soft enough that this does not matter.  Second, one could also allow a homogeneous contribution, representing incoming gravitational waves.  
However, we will shortly pass to an average over a sphere of directions, and this will eliminate any such terms of non-zero helicity.  This would be true whether the homogeneous terms were c-numbers, or multiplied by operators representing linearized gravitational wave creation and annihilation.  So it will apply to any linearized quantum gravity theory in which Einstein's equation holds at an operator level and the homogeneous solutions have helicity two.)

We then find
\begin{eqnarray}\label{deone}
 \frac{d^2\tau _{\rm e}}{d\tau _{\rm r}^2}
 &=&-4\pi G (2\pi )^{-3}m\int_{\|\bk\| ,\, \|\bl\|\leq\Lambda} d^3\bk d^3\bl
 \left[ (2m)^{-2} (2im )^{-1}e^{-2im t}(1-e^{2im D})a(\bk ) a(\bl ) (2m^2)
   +\text{Hermitian conjugate}\right.\nonumber\\
   &&\left.
   +2e^{-i(k_a-l_a)x^a}D
      [(k_a-l_a)(k^a-l^a)]^{-1} a^*(\bl )a(\bk ) 
      (1/2)(\bL\cdot (\bk -\bl ))^2
    \right]\,
\end{eqnarray}
where $D$ is the distance from the emitter to the receiver.
While the first line is simply proportional to the pressure, the second
is rather complicated.  

The most interesting thing about eq. (\ref{deone}) is that integrating it to get $\tau _{\rm e}$ introduces linear terms which must in general be allowed to be operator-valued, for consistency (since additions of constants to $\tau _{\rm r}$ will contribute operator terms to the integral).  Physically, these constants of integration are needed to specify the initial synchronization of the world-lines.  In other words, even in this limited approximation, the effects of quantum geometry on the world-lines' initial data must be considered.
 While this is of considerable conceptual interest, it is better to start with a simpler configuration.

We therefore
imagine a modification of this timing scheme, where instead of having two world-lines, one with a receiver and one with an emitter, we have a single world-line containing an emitter which sends out s-wave pulses, which are then reflected from a sphere of mirrors along worldlines which (at zeroth order) are stationary at distance $D/2$ with respect to the emitter, and are subsequently received on the original world-line.  
(I will discuss the effects of uncertainties in the mirrors' positions below.)
The effect of this on the formula (\ref{deone}) will be to average over the spatial directions $\bL$, and we will get
\begin{eqnarray}\label{detwo}
 \frac{d^2\tau _{\rm e}}{d\tau _{\rm r}^2}
 &=&G(2\pi )^{-2} (4\pi \Lambda ^3/3) \left[
  \sin (mD) e^{-imD}A^2+\text{Hermitian conjugate}
  + (2/3)mD A^*A\right]\nonumber\\
  &=&2\pi G m^{-1}\left[ -2\sin (mD)\overline{P} +(2/3)mD\overline\rho\right]
  \, .
\end{eqnarray}
\end{widetext}

In principle, this is an observable, but it is far too small an effect to be directly detectable in reasonable circumstances.  However, the quantity
\begin{eqnarray}\label{taueq}
\tau _{\rm e}-\tau _{\rm r} &=&D+\\
&&
G(2\pi )^{-2} (4\pi \Lambda ^3/3) \left[\alpha A^2 +{\overline\alpha} (A^*)^2
  +2\beta A^*A\right]\, ,\nonumber
\end{eqnarray}
where
\begin{eqnarray}
  \alpha &=&  -(2m)^{-2}\sin (mD) e^{-imD}\\
  \beta &=&(1/6) mD^3  \, ,
\end{eqnarray}
turns out to be accessible.  (In eq. (\ref{taueq}), the symbol $\overline\alpha$ is the complex conjugate of $\alpha$, not some average.)

The quantity on the square brackets in eq. (\ref{taueq}) can, by a canonical transformation, be rewritten as a number operator if $|\alpha |<|\beta |$ (and as a generator of squeezes if $|\alpha|>|\beta |$).  The condition $|\alpha |<|\beta |$ is equivalent to $mD\gtrsim 1.1$.  Our analysis here is only good in the moderate sector, for which we should have $mD\gtrsim 2\pi$, and thus we are in the number-operator case.  The relevant canonical transformation turns out to be
\begin{eqnarray}
  \hat A &=& (\cosh\xi )    (ie^{-imD/2}A) +(\sinh\xi ) (-ie^{imD/2} A^*)\qquad
\end{eqnarray} 
(so $[\hat A , {\hat A}^*] =1$), with  
\begin{eqnarray}
   \tanh 2\xi &=& e^{imD}\alpha/\beta  =-(3/2)\sin (mD)/(mD)^3\, .
\end{eqnarray}
Then we have   
\begin{eqnarray}
  \alpha A^2 +2\beta A^* A +\overline\alpha (A^*)^2
   &=&2\hat\beta {\hat A}^*\hat A -\hat\varepsilon\, ,
\end{eqnarray}
where 
\begin{eqnarray}
    \hat\beta       
        &=& (1/6) mD^3\sqrt{1+(9/2)(\sin ^2mD)/(mD)^6}
  \\
  \hat\varepsilon      
    &=&(1/6) mD^3\left[ 1-\sqrt{1+(9/2)(\sin ^2mD)/(mD)^6}\right]\, .\qquad
\end{eqnarray}    
In fact, for $mD\geq 2\pi$, we have $\xi \leq 3\times 10^{-4}$, so we have $\hat\beta\approx\beta$, $\hat\alpha \approx 0\approx\hat\varepsilon$.  The quantum correction in eq. (\ref{taueq}) is very nearly simply a multiple of $\overline\rho$, and its analysis parallels that.

For the system under consideration, taking $(4\pi \Lambda ^3/3)A^*A$ to be $(2\pi )^3$ times the number density given by the effective classical theory (see eq. (\ref{rhq})) and $D$ to be a parsec, the magnitude of the quantum contribution in eq. (\ref{taueq}) is $\sim 10^{-7}\, {\rm s}$.

What one would actually measure would be $\tau _{\rm e}-\tau _{\rm r}$.  An observation would yield twice the distance to the reflecting sphere, as the zeroth-order term $D$ plus a quantum-gravitational correction; one cannot distinguish these two terms by this observation.  While in one sense this is disappointing, it does not in fact matter for our main point:

Recall that we are interested, not so much in the value returned by the observation, as in the observation's effect on the quantum state.  For this, we simply need to be able to measure $\tau _{\rm e}-\tau _{\rm r}$ to a resolution fine enough to implicate the quantum corrections, that is, to around $10^{-7}\, {\rm s}$.

To see that there is no difficulty (in principle) in doing this, let me 
return to the question of how well the mirrors' positions must be controlled. 
Since the mirrors will not be exactly spherically distributed, the pulses will return from different mirrors at different times; let us take their mean time of arrival as our $\tau _{\rm e}$.  We then must consider $D$ in eq. (\ref{taueq}) to depend (slightly) on the direction $\bL$.  However, since $\bL$ appears quadratically there, the effect of averaging over the directions will give contributions only from the $l=0$ and $l=2$ multipoles of this function (and their coefficients differ by a factor of order unity).  It is therefore enough to control the quadrupole moments of $D$ to be (say) an order of magnitude smaller than its monopole part, say $\sim 10^{-8}\, \text{l-s}$ or $\sim 1\, {\rm m}$.  This also shows that we need only consider a fairly small number of mirrors, enough to average out the quadrupole.

The requirements imposed by the uncertainty relation on the mirrors' positions and velocities, and on the timing apparatus, are very mild, allowing the devices to be far less massive than the fuzzy dark matter measured, since we only need that the mirrors be controlled to an accuracy of the order of $1\, {\rm m}$ over the time the pulse encounters it.

This argument shows that measurements of $\overline\rho$ do seem possible in principle.

Could one measure the average pressure $\overline{P}$ by a similar procedure?  This does seem possible, although it  is a bit more involved.  To see how one might do this, recall that the operator $A$ is a function of space--time position, and in particular time; in our approximation it evolves with a factor $e^{-imt}$.  In eq. (\ref{taueq}) it is evaluated at the reception point.  If we were to consider a difference in temporal measurements for two pulses, say
\begin{eqnarray}\label{Peek}
&&(\tau _{\rm e}-\tau _{\rm r})\Bigr| _{\tau _{\rm e} =t}
-(\tau _{\rm e}-\tau _{\rm r})\Bigr| _{\tau _{\rm e} =t-\pi /(2m)}\\
&&\qquad= 2G (2\pi )^{-2}(4\pi\Lambda ^3/3)\left[ \alpha A^2 +\overline\alpha (A^*)^2\right]\, ,\nonumber
\end{eqnarray}
this would effect a measurement of ${\overline P}$ (at a time halfway between the two times of receipt).  (For this formula to be directly applicable, one would need a device which reported the double difference (\ref{Peek}), not the two individual $\tau _{\rm e}-\tau _{\rm r}$ measurements.)
The magnitude of this effect would be below that for $\overline\rho$ by a factor of about $\tanh 2\xi$, so it would be $\sim 10^{-11}\, {\rm s}$ or smaller, but there seems to be no objection in principle to measuring it.

\section{Discussion}

Fuzzy dark matter proposals, while generally analyzed classically, are supposed to depend on an underlying quantum field.  Perhaps the most extraordinary feature of this is that its Compton length is of astrophysical size ($\sim 1\, {\rm pc}$); around this scale relativistic quantum effects become important.
In particular, some difficulties in quantum measurement theory are amplified.  

Wigner and followers showed that in general quantum measurements are not compatible with conservation laws; but they also showed that these discrepancies could be made small if the measuring apparatus was much larger than the system measured.  Since this is the case typical in laboratories, and since no violations of fundamental conservation laws have been observed, the problems have not seemed urgent.

However, the large Compton length for fuzzy dark matter allows the ratio between the measuring apparatus and the subject system to be reversed.  In this paper, we have seen it is possible, in principle at least, to have a physical device of modest mass which measures the average of the stress--energy over a few Compton scales (corresponding to $\sim 10^{-2}\, M_\odot$).  We do indeed find serious problems with conservation of energy, if we apply the standard quantum prescriptions.

For the average energy density $\overline\rho$, we find a substantial fraction of the modes in the observation volume become relativistically excited.  This would not only change the energy notably; the resulting particles would escape the Galaxy, and the quantum state would no longer be what was originally hypothesized, a superfluid in a coherent state.

The situation for the average pressure $\overline{P}$ is more difficult to analyze, because it requires us to confront another problem:  the measurement of operators with continuous spectra.  Our treatment focussed on questions of energetics.  There are strong restrictions on the resolutions which could be accommodated by finite-energy measurements.  The simplest natural choices of measurements compatible with those resolutions turn out to lead to quantum states with far higher excitations than for $\overline\rho$.  While this treatment is ad-hoc and cannot be considered definitive, it strongly suggests that $\overline P$ is very singular, insofar as the standard model of quantum measurement is valid.

It does not seem plausible to simply assert that, as we have not so far made any actual measurement of $\overline\rho$ (or $\overline P$), there is not really any problem.  For that defense seems to accept that making these measurements --- which involves only modest actions --- would not only
result in serious violations of energy conservation, but threaten the Galaxy's stability.  This position seems hardly credible.  
Even if one were to accept it, it would lead to another concern:  since the actions required to make a problematic measurement seem modest, one cannot {\em a priori} rule out the possibility that problematic measurements have been effected by physical processes not requiring human intervention in the past few Gy of the Galaxy's history.  This would be a credible concern, and those who wished to simultaneously defend fuzzy dark matter and conventional measurement theory would need to respond to it.  (A response would require answering still another question --- when do measurements occur? --- discussed below.)

There is one assumption in the paper which deserves special attention.  Any attempt to discuss the primary effects of dark matter must turn on its link to gravity, and because here I consider a quantum measurement process it is necessary to relate the geometry of space--time to the quantum stress--energy operator.  I do that by assuming that Einstein's equation holds, at the linearized level, as an equality of operators.  Essentially any attempt to quantize gravity, as that phrase is usually understood (superstring theory, loop quantum gravity, etc.), will have this property.  

Yet this hypothesis could be wrong.  While it is certainly necessary to reconcile gravity and quantum theory, it could be that the way to do so is not by quantizing gravity but by some of other modification.  Perhaps (for example) space--time is inherently classical and so somehow a quantum field must determine a well-defined classical stress--energy. But on its face this would be completely at odds with other quantum-measurement properties of the field.



The simplest possibility would be that somehow fuzzy dark matter is classicized on lengths greater than some scale $\ell \lesssim 1\, {\rm pc}$ (by some new physics).  Then, presumably, the successes of the model in treating galaxies would be secure.  We could still have some sort of quantum-gravitational behavior on scales below $\ell$.  
There would additionally be the question of just how the classicization would affect the particle-physics properties of the field (it is most commonly hypothesized to be an axion), which could be important in the early Universe.

The question of how to modify quantum measurement theory should really be placed in a broader context.
While conventional theory tells us how to model a measurement, given that one occurs, it is entirely silent about when the measurements do occur.  We know of no quantity to compute, for example, which would tell us, given appropriate initial conditions, how likely it is that a given observable will be measured within a given interval of time.  Yet this is an objective question to which physics should be able to supply an answer.

This issue is at the heart of current theories of cosmology:  inflationary models depend on a hypothesized conversion of quantum fluctuations to classical ones, and that appears to be a quantum reduction.  See refs. \cite{PSS2006,MVP2012} for ideas along these lines.

\subsubsection*{Is reduction as a distinct process the problem?}

One could also call into question the reduction postulate for measurements; indeed, there is a spectrum of views on this.  Some physicists accept reduction as it is given in elementary texts; others believe that it ought not to figure in a properly formulated theory, that measurements are just another kind of unitary evolution and it is the idealized treatment of the observed subsystem as an independent entity which forces reduction to appear.  However, there is as yet no developed theory which fully substantiates this belief.  I would suggest that its adherents try to analyze the situation described in this paper, and see if they can improve on the results based on reduction.  
Whatever alternative one envisions, it is essential to have a theory which will correctly describe a sequence of measurements of different observables.

But it is far from clear that reduction, in and of itself, is the culprit.
The problems here arise not precisely from the use of the reduction postulate, but from the identification of the eigenstates of the components of the stress--energy, and their interpretation.  Any substitute for reduction, which has the effect of driving the system into such an eigenstate, will suffer the same problem, unless it provides some further ingredient to resolve that.

Consider, for example, the `relative states formulation' of Everett or the `many worlds interpretation' of DeWitt~\cite{DEG1973}.  These approaches aim to provide a treatment of quantum theory entirely by means of unitary evolution --- but reproducing familiar quantum theory (with reduction) as a sort of `effective' consequence.

I shall not critique these proposals here.  What I want to point out is that, according to their creators, they are supposed to reproduce the results of conventional theory, including reduction.  Their novelty is rather to 
provide additionally what they call a `meta' point of view, and at this `meta' level the reduction is seen to be a short-hand for the creation of certain correlations.  
So if these papers' claims are taken at face value, they will
not resolve the problems uncovered here.  However, both of these papers are rather schematic, and it is possible that a more detailed development of their ideas will uncover structure which will alter this conclusion.

\subsubsection*{de Broglie--Bohm approaches}

Conventional quantum theory is called into question by the results here.  One of the best known alternatives is the de Broglie--Bohm approach~\cite{BH1993,Holland1995,TD2009}.
There remains some controversy over this even at the quantum-mechanical level, and there is no generally accepted way of carrying it over to quantum field theory.  Nevertheless, it has received the attention of serious workers.  I will not critique it, but rather explain what would be involved in attempting to connect it with the work here.

Recall that de Broglie--Bohm theory posits that the positions of particles have real, classical meaning, and that ultimately all measured quantities must derive from these.  There exists also a wave-function satisfying a Schr\"odinger equation, but this does not have at all the same interpretation as in conventional quantum theory.  It rather contributes an additional `quantum force' to the equations of motion of the particles.  Both the trajectories and the wave-functions are classical, although knowledge of them is subject to classical uncertainties.  One should note that the actual position of the system at any time is only a point in the configuration space, while the wave-function is to be defined on that entire space.  As time passes, the actual position describes a curve in configuration space, but the wave-function is function on the Cartesian product of the configuration space with time.  The trajectory is a curve in that space, and so the actual particles only sample directly the values of the wave-function along this curve.\footnote{There is also a `natural' probability distribution one can assign to the positions of the particles, given by the squared modulus of the normalized wave-function; this is called `quantum equilibrium'.  Note however that even if we start from such a distribution, as we learn more about the system we cannot generally maintain it.  Indeed, the equations of motion preserve the `quantum equilibrium' condition, but our increasing knowledge forces the actual probability distribution to become narrower.}

There are three, related, issues of extension involved in developing the theory to the point where it could be applied to the measurement problems in this paper.  The first, already mentioned, is getting a relativistic theory.  Second, one would need to know how de Broglie--Bohm systems act as sources for gravity.  Finally, one needs to know how measurements of space--time geometry feed back on the de Broglie--Bohm system.

One could take the view that in the Galaxy fuzzy dark matter is supposed to be a coherent state of non-relativistic particles, and treat these according to a de Broglie--Bohm prescription.  Insofar as that description reproduces standard non-relativistic quantum theory, one presumably has no trouble.  However, this is not enough to explain how the particles give rise to a source for gravity (does that depend only on the particles, or also on the de Broglie--Bohm wavefunction?  how?).  
Also, while this non-relativistic view may serve as a starting-point for treating dark matter in the Galaxy, one presumably needs a fuller picture to connect with the motivating axion theories, and
with the physics of dark matter in the very early Universe.

I turn now to the question of measurements.  In the de Broglie--Bohm approach, only configuration variables are candidates for measurement, and what one actually observes are macroscopic pointer variables.  One must verify, through modeling of the measurement apparatus, that these are indeed correlated with whatever micro-configuration variables one hopes to learn about.
In the case at hand, I have sketched the construction of a specific device built of fairly simple components (clocks, mirrors, light-sources), and so one has a good template for what a de Broglie--Bohm analysis should try to model.

Pinto-Neto, Santos and Struyve~\cite{PNSS2012,PNSS2014} considered a related problem, not Galactic dark matter, but inflationary fluctuations, and encountered some of these issues.  They used, however, not the particles' positions, but the field values as the configuration space for a de Broglie--Bohm approach.  
They also made certain assumptions about the form of the metric, in terms of the de Broglie--Bohm quantities.  They thus take up, in their context, two of the three issues I mentioned above (the third being the measurement theory of the associated geometry).

\appendix*

\section{Calculations for the squeeze operator}

I work out here some formulas for the squeeze operator
\begin{equation}
  S=(1/2)(A^2 +(A^*)^2)
\end{equation}
figuring in the formula for the pressure, where $A$ is a normalized annihilation operator.  (By a change of phase one can get other squeeze operators.)  The steps are virtually all standard.  I give here the main stages, leaving out some of the straightforward intermediate passages.

Define
\begin{equation}\label{qeq}
  q=2^{-1/2}\left( e^{i\pi /4}A +e^{-i\pi/4}A^*\right)\, ;
\end{equation}
then one can verify  
\begin{equation}
  i\partial/\partial q = 2^{-1/2}i\left( e^{i\pi/4}A-e^{-i\pi/4}A^*\right)
\end{equation}
from the commutation relation  $[A,A^*]=1$.  
Inversely, we have
\begin{equation}\label{Aeq}
  A=2^{-1/2} \left( e^{-i\pi/4}q +e^{-i\pi/4}\partial _q\right)\, ,
\end{equation}
and its Hermitian adjoint.  
Note that these differ from the usual relation between a number representation and a position representation by a rotation by $\pi /4$ in phase space; equivalently, by a square root of the Fourier transform.  Since the Fourier transform is unitary, the norm in the $q$-representation is the standard $L^2$ norm.  Since the vacuum state in the position representation is invariant under Fourier transformation, it retains its standard form $|0\rangle =\pi ^{-1/2}\exp (-q^2/2)$ in the $q$-representation.

The squeeze operator has the simple form
\begin{equation}\label{sqeeq}
  S=-iq\partial _q-i/2
\end{equation}
in terms of $q$.  It evidently generates dilations:  $\exp (i\alpha S) \gamma (q) =e^{-\alpha /2}\gamma (qe^{\alpha})$.  Its eigenvectors are
\begin{equation}\label{qrep}
  \gamma _{\nu,\pm} (q)=(2\pi )^{-1/2}H(\pm q) |q|^{i\nu -1/2}
\end{equation}
(where $H(x)$ is the Heaviside function),
with normalization $\int _{-\infty}^\infty \overline{\gamma _{\nu ,\sigma}} \gamma _{\nu ',\sigma '}\, dq =\delta (\nu -\nu ')\delta _{\sigma ,\sigma'}$.
(The parity eigenstates are of course $2^{-1/2}(\gamma_{\nu,+}\pm\gamma _{\nu ,-})$.)

In this paper, we are interested in understanding the occupation-number content of the eigenstates of $S$.  However, we must explicitly account for the fact that the spectrum of $S$ is continuous, and therefore the operator cannot be measured with infinite precision.  The expectation $\langle \gamma _{\mu ,\pm}|A^*A|\gamma _{\nu ,\pm}\rangle$ does exist in a weak sense, but only when integrated against a restricted, very smooth, class of test functions; this is because states whose $S$-content is restricted to a a sharply demarcated bound necessarily have components with arbitrarily high occupation numbers.

To deal with this, we will consider states which are not quite eigenstates of $S$, but are smeared by Lorentzians:
\begin{equation}
  \xi _{\nu ,\pm} =(2(\Delta\nu )^3/\pi)^{1/2} \int _{-\infty}^\infty \frac{1}{(\acute\nu -\nu )^2+(\Delta\nu) ^2} \gamma _{\acute\nu ,\pm}\, d\acute\nu 
\end{equation}  
is such a state, centered at $\nu$ and width $\Delta\nu$.\footnote{The unusual normalization arises because  here we have a Lorentzian distribution as an {\em amplitude}, rather than a probability.}

A straightforward calculation based on the formulas above gives
\begin{eqnarray}\label{numexp}
&&(4\pi\Lambda ^3/3) \langle\Psi _{\nu\pm}|a^*(\bk )a(\bk )|\Psi _{\nu\pm}\rangle
=N_{\rm cl} -(1/2) +\nonumber\\
&&\qquad\quad
\frac{2(\Delta\nu )^2}{4(\Delta\nu )^2-1}
  +\frac{\nu ^2 +(\Delta \nu )^2 -(3/4)}{2(\Delta\nu )^2-2}(\Delta\nu )^2
  \nonumber\\
  &&\qquad\quad
  -(2\pi )^{3/2} (4\pi\Lambda ^3/3)^{-1/2}\times\nonumber\\
  &&\qquad\quad
  \left(\phi _{\rm cl}e^{i\pi /4}
    \frac{(4\Delta\nu )(1-i\Delta\nu)}{4(\Delta \nu )^2-1}
     +\text{conjugate}\right)\, .\qquad
\end{eqnarray}     
valid for $\Delta\nu >2$.  This is very complicated, but the points we need do not depend on that.  We are primarily interested in the case where $\nu \sim N_{\rm cl}$ is very large.  Then as long as $\Delta\nu /\nu\lesssim 1$, the quantity (\ref{numexp}) is $\sim \nu ^2/2$.


\bibliography{qmfdmprd1.bbl}

\begin{thebibliography}{21}%
\makeatletter
\providecommand \@ifxundefined [1]{%
 \@ifx{#1\undefined}
}%
\providecommand \@ifnum [1]{%
 \ifnum #1\expandafter \@firstoftwo
 \else \expandafter \@secondoftwo
 \fi
}%
\providecommand \@ifx [1]{%
 \ifx #1\expandafter \@firstoftwo
 \else \expandafter \@secondoftwo
 \fi
}%
\providecommand \natexlab [1]{#1}%
\providecommand \enquote  [1]{``#1''}%
\providecommand \bibnamefont  [1]{#1}%
\providecommand \bibfnamefont [1]{#1}%
\providecommand \citenamefont [1]{#1}%
\providecommand \href@noop [0]{\@secondoftwo}%
\providecommand \href [0]{\begingroup \@sanitize@url \@href}%
\providecommand \@href[1]{\@@startlink{#1}\@@href}%
\providecommand \@@href[1]{\endgroup#1\@@endlink}%
\providecommand \@sanitize@url [0]{\catcode `\\12\catcode `\$12\catcode
  `\&12\catcode `\#12\catcode `\^12\catcode `\_12\catcode `\%12\relax}%
\providecommand \@@startlink[1]{}%
\providecommand \@@endlink[0]{}%
\providecommand \url  [0]{\begingroup\@sanitize@url \@url }%
\providecommand \@url [1]{\endgroup\@href {#1}{\urlprefix }}%
\providecommand \urlprefix  [0]{URL }%
\providecommand \Eprint [0]{\href }%
\providecommand \doibase [0]{http://dx.doi.org/}%
\providecommand \selectlanguage [0]{\@gobble}%
\providecommand \bibinfo  [0]{\@secondoftwo}%
\providecommand \bibfield  [0]{\@secondoftwo}%
\providecommand \translation [1]{[#1]}%
\providecommand \BibitemOpen [0]{}%
\providecommand \bibitemStop [0]{}%
\providecommand \bibitemNoStop [0]{.\EOS\space}%
\providecommand \EOS [0]{\spacefactor3000\relax}%
\providecommand \BibitemShut  [1]{\csname bibitem#1\endcsname}%
\let\auto@bib@innerbib\@empty
\bibitem [{\citenamefont {{Hu}}\ \emph {et~al.}(2000)\citenamefont {{Hu}},
  \citenamefont {{Barkana}},\ and\ \citenamefont {{Gruzinov}}}]{HBG2000}%
  \BibitemOpen
  \bibfield  {author} {\bibinfo {author} {\bibfnamefont {W.}~\bibnamefont
  {{Hu}}}, \bibinfo {author} {\bibfnamefont {R.}~\bibnamefont {{Barkana}}}, \
  and\ \bibinfo {author} {\bibfnamefont {A.}~\bibnamefont {{Gruzinov}}},\
  }\href {\doibase 10.1103/PhysRevLett.85.1158} {\bibfield  {journal} {\bibinfo
   {journal} {Physical Review Letters}\ }\textbf {\bibinfo {volume} {85}},\
  \bibinfo {pages} {1158} (\bibinfo {year} {2000})},\ \Eprint
  {http://arxiv.org/abs/astro-ph/0003365} {astro-ph/0003365} \BibitemShut
  {NoStop}%
\bibitem [{\citenamefont {{Hui}}\ \emph {et~al.}(2017)\citenamefont {{Hui}},
  \citenamefont {{Ostriker}}, \citenamefont {{Tremaine}},\ and\ \citenamefont
  {{Witten}}}]{HOTW2017}%
  \BibitemOpen
  \bibfield  {author} {\bibinfo {author} {\bibfnamefont {L.}~\bibnamefont
  {{Hui}}}, \bibinfo {author} {\bibfnamefont {J.~P.}\ \bibnamefont
  {{Ostriker}}}, \bibinfo {author} {\bibfnamefont {S.}~\bibnamefont
  {{Tremaine}}}, \ and\ \bibinfo {author} {\bibfnamefont {E.}~\bibnamefont
  {{Witten}}},\ }\href {\doibase 10.1103/PhysRevD.95.043541} {\bibfield
  {journal} {\bibinfo  {journal} {Phys. Rev. D}\ }\textbf {\bibinfo {volume}
  {95}},\ \bibinfo {eid} {043541} (\bibinfo {year} {2017})},\ \Eprint
  {http://arxiv.org/abs/1610.08297} {arXiv:1610.08297} \BibitemShut {NoStop}%
\bibitem [{\citenamefont {Wigner}(1952)}]{Wigner1952}%
  \BibitemOpen
  \bibfield  {author} {\bibinfo {author} {\bibfnamefont {E.~P.}\ \bibnamefont
  {Wigner}},\ }\href@noop {} {\bibfield  {journal} {\bibinfo  {journal}
  {Zeitschrift f\"ur Physik}\ }\textbf {\bibinfo {volume} {133}},\ \bibinfo
  {pages} {101} (\bibinfo {year} {1952})}\BibitemShut {NoStop}%
\bibitem [{\citenamefont {{Ahmadi}}\ \emph {et~al.}(2013)\citenamefont
  {{Ahmadi}}, \citenamefont {{Jennings}},\ and\ \citenamefont
  {{Rudolph}}}]{AJR}%
  \BibitemOpen
  \bibfield  {author} {\bibinfo {author} {\bibfnamefont {M.}~\bibnamefont
  {{Ahmadi}}}, \bibinfo {author} {\bibfnamefont {D.}~\bibnamefont
  {{Jennings}}}, \ and\ \bibinfo {author} {\bibfnamefont {T.}~\bibnamefont
  {{Rudolph}}},\ }\href {\doibase 10.1088/1367-2630/15/1/013057} {\bibfield
  {journal} {\bibinfo  {journal} {New Journal of Physics}\ }\textbf {\bibinfo
  {volume} {15}},\ \bibinfo {eid} {013057} (\bibinfo {year} {2013})},\ \Eprint
  {http://arxiv.org/abs/1209.0921} {arXiv:1209.0921 [quant-ph]} \BibitemShut
  {NoStop}%
\bibitem [{\citenamefont {Leggett}(2002)}]{Leggett2002}%
  \BibitemOpen
  \bibfield  {author} {\bibinfo {author} {\bibfnamefont {A.~J.}\ \bibnamefont
  {Leggett}},\ }\href {http://stacks.iop.org/0953-8984/14/i=15/a=201}
  {\bibfield  {journal} {\bibinfo  {journal} {Journal of Physics: Condensed
  Matter}\ }\textbf {\bibinfo {volume} {14}},\ \bibinfo {pages} {R415}
  (\bibinfo {year} {2002})}\BibitemShut {NoStop}%
\bibitem [{\citenamefont {{Khmelnitsky}}\ and\ \citenamefont
  {{Rubakov}}(2014)}]{KR2014}%
  \BibitemOpen
  \bibfield  {author} {\bibinfo {author} {\bibfnamefont {A.}~\bibnamefont
  {{Khmelnitsky}}}\ and\ \bibinfo {author} {\bibfnamefont {V.}~\bibnamefont
  {{Rubakov}}},\ }\href {\doibase 10.1088/1475-7516/2014/02/019} {\bibfield
  {journal} {\bibinfo  {journal} {Journal of Cosmology and Astroparticle
  Physics}\ }\textbf {\bibinfo {volume} {2}},\ \bibinfo {eid} {019} (\bibinfo
  {year} {2014})},\ \Eprint {http://arxiv.org/abs/1309.5888} {arXiv:1309.5888}
  \BibitemShut {NoStop}%
\bibitem [{\citenamefont {von Neumann}(1955)}]{vN1955}%
  \BibitemOpen
  \bibfield  {author} {\bibinfo {author} {\bibfnamefont {J.}~\bibnamefont {von
  Neumann}},\ }\href@noop {} {\emph {\bibinfo {title} {Mathematical foundations
  of quantum mechanics}}},\ \bibinfo {series} {Investigations in Physics}\
  No.~\bibinfo {number} {2}\ (\bibinfo  {publisher} {Princeton University
  Press},\ \bibinfo {year} {1955})\BibitemShut {NoStop}%
\bibitem [{\citenamefont {Helfer}(2004)}]{ADH2004a}%
  \BibitemOpen
  \bibfield  {author} {\bibinfo {author} {\bibfnamefont {A.~D.}\ \bibnamefont
  {Helfer}},\ }\href@noop {} {\bibfield  {journal} {\bibinfo  {journal}
  {Physics Letters}\ }\textbf {\bibinfo {volume} {A329}},\ \bibinfo {pages}
  {277} (\bibinfo {year} {2004})}\BibitemShut {NoStop}%
\bibitem [{\citenamefont {Helfer}(2012)}]{ADH2011}%
  \BibitemOpen
  \bibfield  {author} {\bibinfo {author} {\bibfnamefont {A.~D.}\ \bibnamefont
  {Helfer}},\ }in\ \href@noop {} {\emph {\bibinfo {booktitle} {Cosmology and
  Gravitation: XIVth Brazilian School of Cosmology and Gravitation}}},\
  \bibinfo {editor} {edited by\ \bibinfo {editor} {\bibfnamefont
  {M.}~\bibnamefont {{Novello}}}\ and\ \bibinfo {editor} {\bibfnamefont
  {S.~E.}\ \bibnamefont {{Perez~Bergliaffa}}}}\ (\bibinfo  {publisher}
  {Cambridge Scientific},\ \bibinfo {year} {2012})\ pp.\ \bibinfo {pages}
  {45--81},\ \Eprint {http://arxiv.org/abs/1105.1980} {arXiv:1105.1980 [gr-qc]}
  \BibitemShut {NoStop}%
\bibitem [{\citenamefont {{Bedingham}}\ \emph {et~al.}(2016)\citenamefont
  {{Bedingham}}, \citenamefont {{Modak}},\ and\ \citenamefont
  {{Sudarsky}}}]{BMS2016}%
  \BibitemOpen
  \bibfield  {author} {\bibinfo {author} {\bibfnamefont {D.}~\bibnamefont
  {{Bedingham}}}, \bibinfo {author} {\bibfnamefont {S.~K.}\ \bibnamefont
  {{Modak}}}, \ and\ \bibinfo {author} {\bibfnamefont {D.}~\bibnamefont
  {{Sudarsky}}},\ }\href {\doibase 10.1103/PhysRevD.94.045009} {\bibfield
  {journal} {\bibinfo  {journal} {Phys. Rev. D}\ }\textbf {\bibinfo {volume}
  {94}},\ \bibinfo {eid} {045009} (\bibinfo {year} {2016})},\ \Eprint
  {http://arxiv.org/abs/1604.06537} {arXiv:1604.06537 [gr-qc]} \BibitemShut
  {NoStop}%
\bibitem [{\citenamefont {Schweber}(1961)}]{Schweber1961}%
  \BibitemOpen
  \bibfield  {author} {\bibinfo {author} {\bibfnamefont {S.~S.}\ \bibnamefont
  {Schweber}},\ }\href@noop {} {\emph {\bibinfo {title} {An Introduction to
  Relativistic Quantum Field Theory}}}\ (\bibinfo  {publisher} {Row, Peterson
  and Company},\ \bibinfo {year} {1961})\BibitemShut {NoStop}%
\bibitem [{\citenamefont {Penrose}\ and\ \citenamefont
  {Rindler}(1984)}]{PR1984}%
  \BibitemOpen
  \bibfield  {author} {\bibinfo {author} {\bibfnamefont {R.}~\bibnamefont
  {Penrose}}\ and\ \bibinfo {author} {\bibfnamefont {W.}~\bibnamefont
  {Rindler}},\ }\href@noop {} {\emph {\bibinfo {title} {Spinors and
  space--time, vol. 1: Two--spinor calculus and relativistic fields}}}\
  (\bibinfo  {publisher} {Cambridge University Press},\ \bibinfo {year}
  {1984})\BibitemShut {NoStop}%
\bibitem [{\citenamefont {{Helfer}}(2013)}]{ADH2013}%
  \BibitemOpen
  \bibfield  {author} {\bibinfo {author} {\bibfnamefont {A.~D.}\ \bibnamefont
  {{Helfer}}},\ }\href {\doibase 10.1093/mnras/sts618} {\bibfield  {journal}
  {\bibinfo  {journal} {Mon. Not. R. Astron. Soc.}\ }\textbf {\bibinfo {volume}
  {430}},\ \bibinfo {pages} {305} (\bibinfo {year} {2013})},\ \Eprint
  {http://arxiv.org/abs/1212.2926} {arXiv:1212.2926 [gr-qc]} \BibitemShut
  {NoStop}%
\bibitem [{\citenamefont {{Perez}}\ \emph {et~al.}(2006)\citenamefont
  {{Perez}}, \citenamefont {{Sahlmann}},\ and\ \citenamefont
  {{Sudarsky}}}]{PSS2006}%
  \BibitemOpen
  \bibfield  {author} {\bibinfo {author} {\bibfnamefont {A.}~\bibnamefont
  {{Perez}}}, \bibinfo {author} {\bibfnamefont {H.}~\bibnamefont {{Sahlmann}}},
  \ and\ \bibinfo {author} {\bibfnamefont {D.}~\bibnamefont {{Sudarsky}}},\
  }\href {\doibase 10.1088/0264-9381/23/7/008} {\bibfield  {journal} {\bibinfo
  {journal} {Classical and Quantum Gravity}\ }\textbf {\bibinfo {volume}
  {23}},\ \bibinfo {pages} {2317} (\bibinfo {year} {2006})},\ \Eprint
  {http://arxiv.org/abs/gr-qc/0508100} {gr-qc/0508100} \BibitemShut {NoStop}%
\bibitem [{\citenamefont {{Martin}}\ \emph {et~al.}(2012)\citenamefont
  {{Martin}}, \citenamefont {{Vennin}},\ and\ \citenamefont
  {{Peter}}}]{MVP2012}%
  \BibitemOpen
  \bibfield  {author} {\bibinfo {author} {\bibfnamefont {J.}~\bibnamefont
  {{Martin}}}, \bibinfo {author} {\bibfnamefont {V.}~\bibnamefont {{Vennin}}},
  \ and\ \bibinfo {author} {\bibfnamefont {P.}~\bibnamefont {{Peter}}},\ }\href
  {\doibase 10.1103/PhysRevD.86.103524} {\bibfield  {journal} {\bibinfo
  {journal} {Phys. Rev. D}\ }\textbf {\bibinfo {volume} {86}},\ \bibinfo {eid}
  {103524} (\bibinfo {year} {2012})},\ \Eprint {http://arxiv.org/abs/1207.2086}
  {arXiv:1207.2086 [hep-th]} \BibitemShut {NoStop}%
\bibitem [{\citenamefont {{Dewitt}}\ \emph {et~al.}(1973)\citenamefont
  {{Dewitt}}, \citenamefont {{Everett}},\ and\ \citenamefont
  {{Graham}}}]{DEG1973}%
  \BibitemOpen
  \bibfield  {author} {\bibinfo {author} {\bibfnamefont {B.~S.}\ \bibnamefont
  {{Dewitt}}}, \bibinfo {author} {\bibfnamefont {H.}~\bibnamefont {{Everett}}},
  \ and\ \bibinfo {author} {\bibfnamefont {N.}~\bibnamefont {{Graham}}},\
  }\href@noop {} {\emph {\bibinfo {title} {The many-worlds interpretation of
  quantum mechanics}}},\ Princeton Series in Physics\ (\bibinfo  {publisher}
  {Princeton University Press},\ \bibinfo {address} {Princeton, N.J.},\
  \bibinfo {year} {1973})\BibitemShut {NoStop}%
\bibitem [{\citenamefont {{Bohm}}\ and\ \citenamefont
  {{Hiley}}(1993)}]{BH1993}%
  \BibitemOpen
  \bibfield  {author} {\bibinfo {author} {\bibfnamefont {D.}~\bibnamefont
  {{Bohm}}}\ and\ \bibinfo {author} {\bibfnamefont {B.~J.}\ \bibnamefont
  {{Hiley}}},\ }\href@noop {} {\emph {\bibinfo {title} {{The undivided
  universe. an ontological interpretation of quantum theory}}}}\ (\bibinfo
  {publisher} {Routledge},\ \bibinfo {address} {London, New York},\ \bibinfo
  {year} {1993})\BibitemShut {NoStop}%
\bibitem [{\citenamefont {{Holland}}(1995)}]{Holland1995}%
  \BibitemOpen
  \bibfield  {author} {\bibinfo {author} {\bibfnamefont {P.~R.}\ \bibnamefont
  {{Holland}}},\ }\href@noop {} {\emph {\bibinfo {title} {{The Quantum Theory
  of Motion}}}}\ (\bibinfo  {publisher} {Cambridge University Press},\ \bibinfo
  {address} {Cambridge, U.K.},\ \bibinfo {year} {1995})\ p.\ \bibinfo {pages}
  {618}\BibitemShut {NoStop}%
\bibitem [{\citenamefont {{Teufel}}\ and\ \citenamefont
  {{D{\"u}rr}}(2009)}]{TD2009}%
  \BibitemOpen
  \bibfield  {author} {\bibinfo {author} {\bibfnamefont {S.}~\bibnamefont
  {{Teufel}}}\ and\ \bibinfo {author} {\bibfnamefont {D.}~\bibnamefont
  {{D{\"u}rr}}},\ }\href {\doibase 10.1007/b99978} {\emph {\bibinfo {title}
  {{Bohmian Mechanics}}}}\ (\bibinfo  {publisher} {Springer-Verlag},\ \bibinfo
  {address} {Berline, Heidelberg},\ \bibinfo {year} {2009})\BibitemShut
  {NoStop}%
\bibitem [{\citenamefont {{Pinto-Neto}}\ \emph {et~al.}(2012)\citenamefont
  {{Pinto-Neto}}, \citenamefont {{Santos}},\ and\ \citenamefont
  {{Struyve}}}]{PNSS2012}%
  \BibitemOpen
  \bibfield  {author} {\bibinfo {author} {\bibfnamefont {N.}~\bibnamefont
  {{Pinto-Neto}}}, \bibinfo {author} {\bibfnamefont {G.}~\bibnamefont
  {{Santos}}}, \ and\ \bibinfo {author} {\bibfnamefont {W.}~\bibnamefont
  {{Struyve}}},\ }\href {\doibase 10.1103/PhysRevD.85.083506} {\bibfield
  {journal} {\bibinfo  {journal} {\prd}\ }\textbf {\bibinfo {volume} {85}},\
  \bibinfo {eid} {083506} (\bibinfo {year} {2012})},\ \Eprint
  {http://arxiv.org/abs/1110.1339} {arXiv:1110.1339 [gr-qc]} \BibitemShut
  {NoStop}%
\bibitem [{\citenamefont {{Pinto-Neto}}\ \emph {et~al.}(2014)\citenamefont
  {{Pinto-Neto}}, \citenamefont {{Santos}},\ and\ \citenamefont
  {{Struyve}}}]{PNSS2014}%
  \BibitemOpen
  \bibfield  {author} {\bibinfo {author} {\bibfnamefont {N.}~\bibnamefont
  {{Pinto-Neto}}}, \bibinfo {author} {\bibfnamefont {G.}~\bibnamefont
  {{Santos}}}, \ and\ \bibinfo {author} {\bibfnamefont {W.}~\bibnamefont
  {{Struyve}}},\ }\href {\doibase 10.1103/PhysRevD.89.023517} {\bibfield
  {journal} {\bibinfo  {journal} {\prd}\ }\textbf {\bibinfo {volume} {89}},\
  \bibinfo {eid} {023517} (\bibinfo {year} {2014})},\ \Eprint
  {http://arxiv.org/abs/1309.2670} {arXiv:1309.2670 [gr-qc]} \BibitemShut
  {NoStop}%
\end{thebibliography}%

\end{document}